\documentclass[twocolumn]{aastex62}
\shorttitle{First Detection of Radio Linear Polarization in a Gamma Ray Burst Afterglow}
\shortauthors{Urata, Toma et al.}
\begin{document}

\title{First Detection of Radio Linear Polarization in a Gamma Ray Burst Afterglow}

\correspondingauthor{Yuji Urata}
\email{urata@g.ncu.edu.tw}

\author{Yuji Urata}
\affiliation{Institute of Astronomy, National Central University, Chung-Li 32054, Taiwan}

\author{Kenji Toma}
\affiliation{Frontier Research Institute for Interdisciplinary Sciences, Tohoku University, Sendai 980-8578, Japan}
\affiliation{Astronomical Institute, Tohoku University, Sendai, 980- 8578, Japan}

\author{Kuiyun Huang}
\affiliation{Center for General Education, Chung Yuan Christian University, Taoyuan 32023, Taiwan}

\author{Keiichi Asada}
\affiliation{Academia Sinica Institute of Astronomy and Astrophysics, Taipei 106, Taiwan}

\author{Hiroshi Nagai}
\affiliation{National Astronomical Observatory of Japan, 2-21-1 Osawa, Mitaka Tokyo 181-8588, Japan}
\affiliation{Department of Astronomical Science, School of Physical Sciences, SOKENDAI (The Graduate University for Advanced Studies), Mitaka, Tokyo 181-8588, Japan}

\author{Satoko Takahashi}
\affiliation{Joint ALMA Observatory, Alonso de Cordova 3108, Vitacura, Santiago, Chile}
\affiliation{NAOJ Chile Observatory, Alonso de Cordova 3788, Oficina 61B, Vitacura, Santiago, Chile}
\affiliation{Department of Astronomical Science, School of Physical Sciences, SOKENDAI (The Graduate University for Advanced Studies), Mitaka, Tokyo 181-8588, Japan}

\author{Glen Petitpas}
\affiliation{Harvard-Smithsonian Center for Astrophysics, 60 Garden Street, Cambridge, Massachusetts 02138, USA}

\author{Makoto Tashiro}
\affiliation{Department of Physics, Saitama University, Shimo-Okubo, Saitama, 338-8570, Japan}

\author{Kazutaka Yamaoka}
\affiliation{Institute for Space-Earth Environmental Research (ISEE), Nagoya University, Furo-cho, Chikusa-ku, Nagoya, Aichi 464- 8601, Japan}
\affiliation{Division of Particle and Astrophysical Science, Graduate School of Science, Nagoya University, Furo-cho, Chikusa-ku, Nagoya, Aichi 464-8601, Japan}

%\author{}
%\affiliation{}

%% Mark off the abstract in the ``abstract'' environment. 
\begin{abstract}

We report the first detection of radio polarization of a GRB afterglow
with the first intensive combined use of telescopes in the millimeter
and submillimeter ranges for GRB171205A.  The linear polarization
degree in the millimeter band at the sub-percent level ($0.27 \pm
0.04\%$) is lower than those observed in late-time optical afterglows
(weighted average of $\sim 1\%$). The Faraday depolarization by
non-accelerated, cool electrons in the shocked region is one of
possible mechanisms for the low value. In this scenario, larger total
energy by a factor of $\sim 10$ than ordinary estimate without
considering non-accelerated electrons is required.  The polarization
position angle varies by at least 20 degrees across the millimeter
band, which is not inconsistent with this scenario.  This result
indicates that polarimetry in the millimeter and submillimeter ranges
is a unique tool for investigating GRB energetics, and coincident
observations with multiple frequencies or bands would provide more
accurate measurements of the non-accelerated electron fraction.

\end{abstract}
%
%% Keywords should appear after the \end{abstract} command. 
%% See the online documentation for the full list of available subject
%% keywords and the rules for their use.

%% From the front matter, we move on to the body of the paper.
%% Sections are demarcated by \section and \subsection, respectively.
%% Observe the use of the LaTeX \label
%% command after the \subsection to give a symbolic KEY to the
%% subsection for cross-referencing in a \ref command.
%% You can use LaTeX's \ref and \label commands to keep track of
%% cross-references to sections, equations, tables, and figures.
%% That way, if you change the order of any elements, LaTeX will
%% automatically renumber them.
%%
%% We recommend that authors also use the natbib \citep
%% and \citet commands to identify citations.  The citations are
%% tied to the reference list via symbolic KEYs. The KEY corresponds
%% to the KEY in the \bibitem in the reference list below. 

\section{Introduction} \label{sec:intro}

Gamma-ray Bursts (GRBs) are highly energetic explosions in the
universe, and are currently being exploited as probes of
first-generation stars and gravitational wave transients. In fact, the
distant events at the re-ionization epoch \citep{1,2,3} and the short
GRB coincident with a gravitational wave transient have already been
observed \citep{4}, respectively. The energetics of GRBs are
fundamental physical parameters that can not only reveal their
progenitor systems but also probe both the early and current states of
the universe\citep[e.g.,][]{murase06,toma16,kawaguchi18}. Although
substantial observational efforts have been made since the afterglow
discovery \citep{5}, the total energies have been estimated so far
without considering non-accelerated, cool electrons at the
relativistic collisionless shocks that do not emit observable
radiation \citep{6}, while the existence of such cool electrons is
well studied for supernova remnants and solar winds
\citep[e.g.][]{7,8}. 
In GRB afterglows, the presence of non-accelerated electrons would induce Faraday effects on the emitted radiation. Observationally, this manifests as a suppression of the radio polarization but keeps the optical polarization as emitted \citep{9}\footnote{Spectroscopic searches of non-accelerated electrons are discussed in \cite{35} and \cite{36}.}. 
Here, we report the first detection of radio polarization of a GRB
afterglow through observing low-luminosity GRB 171205A, and discuss
implications for the Faraday depolarization model.

GRB 171205A was detected on 5 December 2017, 07:20:43 UT \citep{13} and
its X-ray and optical afterglows \citep{14} are identified by the Neil
Gehrels Swift Observatory. \citet{15} made spectroscopic observations
with the Very Large Telescope (VLT) in Chile approximately 1.5 h after
the GRB by identifying the optical afterglow and, based on the
absorption and emission lines, announced a redshift of $z =$ 0.0368.
At this redshift, the
isotropic $\gamma$-ray energy release $E_{\gamma,{\rm iso}}$ of
$2.4\times10^{49}~$erg (in the 20$-$1500 keV range with the cosmological
parameters $H_{0}=$ 70 ${\rm km}$ ${\rm s}^{-1}$ ${\rm Mpc}^{-1}$,
$\Omega_m=0.3$, and $\Omega_{\Lambda}=0.7$) indicates that
GRB171205A is categorized as a low-luminosity GRB. Intensive optical
photometric and spectroscopic observations using the 10.4-m Gran
Telescopio CANARIAS (GTC) revealed the association of a broad-line
type Ic supernova that resembled SN1998bw \citep{16}. The bright
millimeter afterglow was also detected by the Northern Extended
Millimeter Array (NOEMA) in the 90 GHz and 150 GHz bands 20.2 h after
the burst \citep{17}.

\section{Observations and Analysis} \label{sec:obs}

\subsection{SMA}

Intensive total flux monitoring was made using the
Submillimeter Array (SMA) at 230 GHz starting 6 December 2017 with a
total of six epochs. On the nights of 8 and 13 December 2017, the
afterglow was observed by the dual-band mode at 230 and 345 GHz. The
data were flagged and calibrated with the MIR data-reduction package
using standard procedures and were then imaged using Miriad software
\citep{31}. Except for the observation at 345 GHz on 13 December 2017
(due to marginal weather conditions in the band), the afterglow was
clearly detected at a confidence level of more than 10$\sigma$. Flux
measurements were performed using Common Astronomy Software
Applications \citep[CASA, version 5.1.1;][]{32}.
%
%We observed the afterglow position with the Submillimetre Array (SMA)
%located on Mauna Kea in Hawaii, starting on the night of 6 December
%2017.
We measured a bright submillimeter afterglow of $53.7 \pm 0.9~$mJy in
the 230 GHz band 1.5 days after GRB, which is the brightest afterglow
ever detected in the submillimetre range. At the same epoch, the
historic GRB030329 was $49.2 \pm 1.1~$mJy in the 250 GHz band
\citep{18}, while typical submm-detected afterglows are orders of
magnitude fainter\citep{19,20}.  Thus, GRB 171205A is an ideal object
for performing the first radio polarimetry.

\begin{figure}
\epsscale{1.2}
\plotone{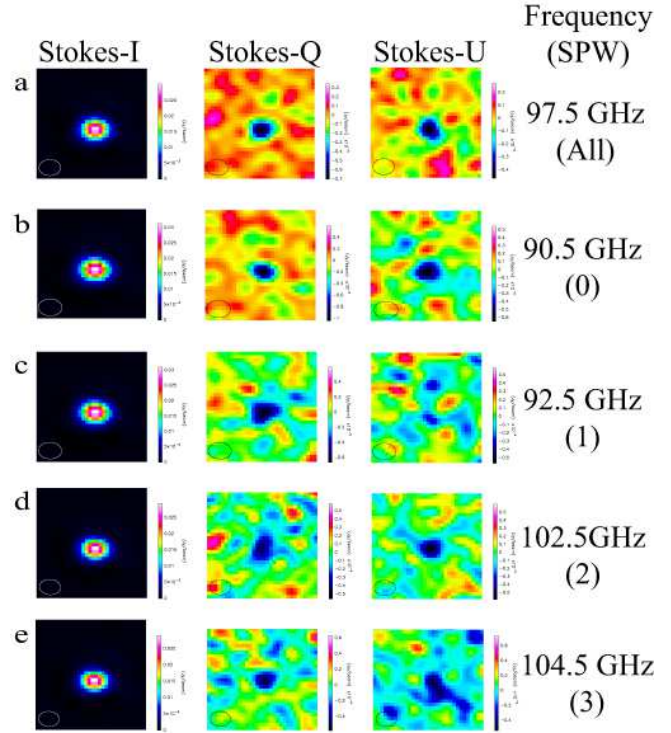}
\caption{The Stokes $I, Q,$ and $U$ maps (5$''\times3''$) of the afterglow of GRB 171205A taken on 10 December 2018 (5.187 days after the burst). The ALMA beam size is shown with the open cyan circles. The map created using the entire ALMA Band 3 dataset with a representative frequency of 97.5 GHz (a), and four individual spectral windows (SPW) with a representative frequency of 90.5 GHz (b), 92.5 GHz (c), 102.5 GHz (d), and 104.5 GHz (e).  The units of color bars are mJy for Stokes $I$ and $\mu$Jy for Stokes $Q$ and $U$ maps.
}
\label{polimage}
\end{figure}
\subsection{ALMA}

The Atacama Large Millimeter/submillimeter Array (ALMA) observed the
afterglow in two different epochs using the linear polarization mode
at Band 3 (representative frequency of 97.5 GHz) on 10 and 16 December
2017. The correlator processed four spectral windows (SPWs) centered
at 90.5, 92.5, 102.5, and 104.5 GHz with a bandwidth of 1.75 GHz
each. The bandpass and flux were calibrated using observations of
J1127-1857, and J1130-1149 was used for the phase calibration. The
polarization calibration was performed by observations of
J1256-0547. The raw data were reduced at the East Asian ALMA Regional
Center (EA-ARC) using CASA (version 5.1.1). We further performed
interative CLEAN deconvolution imaging with self-calibration both
amplitude and phase with infinite and then 30s solution
intervals). The Stokes $I, Q,$ and $U$ maps were CLEANed
up to 15000 of CLEAN iterations with threshold of 0.02 mJy after the final round
of self-calibration (Figure \ref{polimage}a).
The off-source rms levels in $I$, $Q$, and $U$ are consistent with the expectations for thermal noise alone.
Since the detections with high signal to noise ratio were made on the Stokes $Q$ and $U$ maps
generated using the entire Band 3 dataset from 10 December 2017, we
generated additional Stokes maps using the individual SPWs (Figure \ref{polimage}b,c,d,e).
The quantities that can be derived from the polarization maps are the
polarized intensity ($\sqrt{Q^2+U^2}$), polarization
degree (100$\sqrt{Q^2+U^2}/I$\%), and polarization position angle
(1/2$\arctan(U/Q)$, P.A.). The $atan2$ function
in the python math module which returns a numeric value between $-\pi$
and $\pi$, was used to calculate the polarization position angle.
By applying the polarization calibration to the phase
  calibrator J1130-1449 and creating Stokes maps for 6, 9, and 18 epochs during
  the 3hr of observing period, we confirm that the stability of linear
  polarization degree is $<0.02$\%, which is consistent with the
  systematic linear polarization calibration uncertainty of 0.033\%
  for compact sources\footnote{ALMA technical
    handbook;https://almascience.nrao.edu/documents-and-tools/cycle7/alma-technical-handbook/}. We
  also find that the stability of P.A. is $<0^\circ.6$, which is
  slightly larger than the absolute accuracy of
  $0.3^{\circ}$\citep{nagai}. 
The non-detection (both positive and negative) with $S/N$ of 3 on the
92.5-GHz $U$ map taken on 10 December 2017 yielded polarization
position angle ranges of P.A. $>+78^{\circ}$ and P.A. $<-78^{\circ}$.

The Atacama Compact Array (ACA) observations were executed on 10, 12,
and 16 December 2017 at 345 GHz (Band 7) with the single continuum
observing mode. Two of the ACA total flux measurments were conducted
during polarimetry using ALMA. The data were flagged, calibrated and
imaged with standard procedures with CASA (version 5.1.1).

%We conducted two epochs of radio linear polarimetry using the Atacama
%Large Millimetre/submillimetre Array (ALMA) with the suitably high
%sensitivity for polarimetry in Chile at 97.5 GHz (i.e. Band 3). The
%observations were also managed together with photometry at 343.5 GHz
%using the Atacama Compact Array (ACA) at the same observing
%site. Additional five epochs of 230 GHz photometric monitoring were
%performed using SMA until 11 days after the burst.

\begin{figure}
\epsscale{1.20}
\plotone{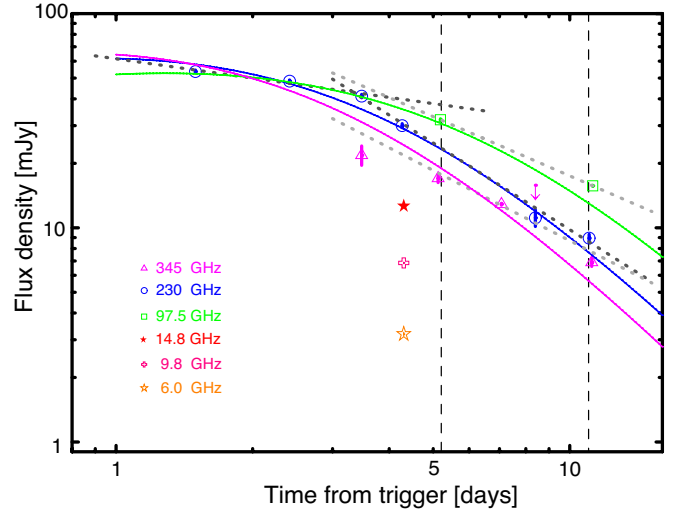}
\caption{Radio afterglow light curves. Solid lines indicate the model light curves at 97.5 GHz (green), 230 GHz (blue), and 345 GHz (magenta) based on the standard forward shock model. Dark grey dotted lines show the simple power-law fittings for 230 GHz data before and 4 days after the burst. Light grey dotted lines show the simple power-law functions for 97.5 GHz with $\alpha=-0.9$ and 345 GHz with $\alpha=-1.2$. Thin black dashed lines indicate the epochs of ALMA polarimetry.
}
\label{lc}
\end{figure}

\subsection{VLA}

The Very Large Array (VLA) made total flux measurements for the
afterglow on 9 December 2017 at central frequencies of 6 GHz (C-band),
10 GHz (X-band), and 15 GHz (U-band), as one of the
observatory-sponsored observations \citep{21}. The phase and flux were
calibrated using observations of J1130-1449 and 3C286. The data were
calibrated using standard tools in CASA (VLA pipeline version
5.0.0). After checking the quality of the pipeline output,
  we performed imaging using CLEAN task without additional data
  flagging. The source was significantly (more than 50$\sigma$)
detected in all three bands. To describe the spectral energy
distribution, six images at the central frequencies of 5 GHz, 7 GHz,
8.5 GHz, 11 GHz, 13.5 GHz, and 16 GHz were generated with the CLEAN
task. The afterglow was detected with $>$20$\sigma$
  significance in each image and the resulting total flux densities
  are summarized in Table \ref{almapol}.

\begin{figure*}
\epsscale{.7}
\plotone{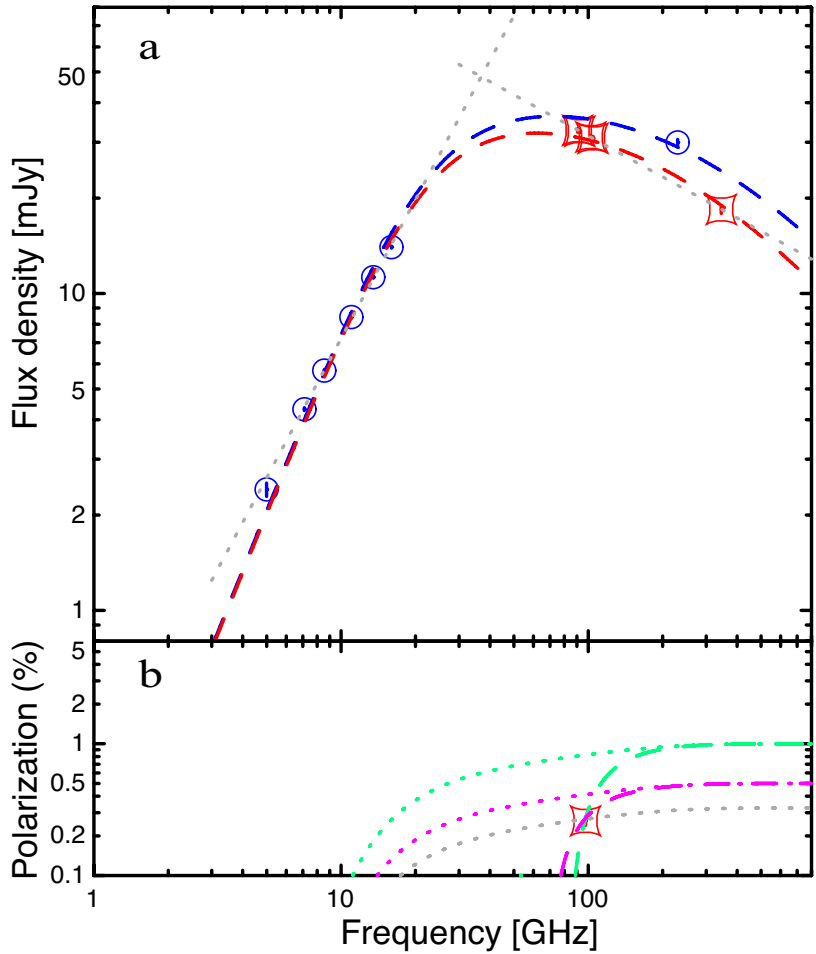}
\caption{Spectral flux distributions and total linear polarization spectrum of the GRB 171205A afterglow. a, Spectral flux distribution at 4.1 days (blue circles and model dashed line) and 5.2 days (red squared points and model dashed line) after burst. The grey dotted lines indicate the simple power-law functions with index of 1.457 and -0.430. b, Total linear polarization spectrum with the ALMA measurement 5.2 days after the burst. Dashed lines indicate the Faraday depolarized spectrum by assuming $P_{0}$ of 1\% (green) and 0.5\% (magenta). The dotted lines indicate the polarization spectrum without the Faraday depolarization effect (i.e. all electrons are energized by the relativistic shock) by assuming $P_{0}$ of 1\% (green), 0.5\% (magenta), and 0.33\% (grey).}
\label{polspec}
\end{figure*}

\section{Results} \label{sec:result}

\subsection{Lightcurve and SED}

The temporal evolution of the afterglow flux at 230 GHz is described
by broken power-law decays ($F_{\nu} \propto t^{\alpha}$) with $\alpha
= -0.30 \pm 0.07$ for $t\lesssim4$ days and $\alpha = -1.34 \pm 0.06$
for $t\gtrsim4$ days, as show in Figure \ref{lc}. The light curve at
345 GHz for $t\gtrsim4$ is also described by a simple power-law with
$\alpha = -1.2 \pm 0.2$.
The spectral slope ($F_{\nu} \propto \nu^{\beta}$) is also described
as $\beta =1.457 \pm 0.028$ at 4.3 days in the centimeter range
(5$-$16 GHz; Figure \ref{polspec}a) and $\beta=-0.430 \pm 0.004$ at
5.2 days in the submillimeter and millimeter range (90.5$-$345 GHz;
Figure \ref{polspec}a). High-quality photometry ($S/N \sim$ 72$-$89)
using ALMA during the polarimetry, at 5.2 days, measured the spectral
slope of $\beta = -0.40 \pm 0.01$ in the 90$-$100 GHz (i.e. Band
3). These measurements indicate that the spectral peak was located at
$\sim 30~$GHz (below $\sim 90~$GHz).

\begin{figure*}
\epsscale{0.7}
\plotone{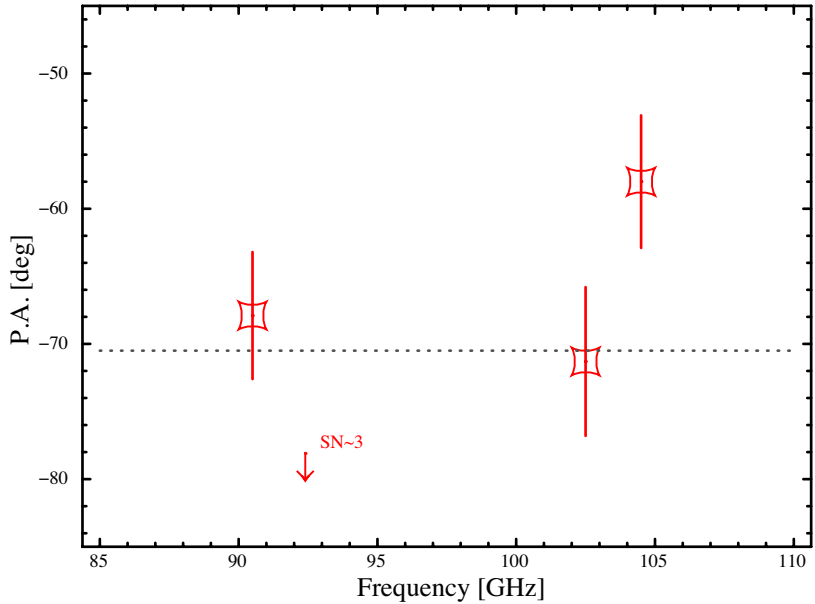}
\caption{The position angle (P.A.) of the GRB171205A afterglow as a function of wavelength. Squared points indicate the observed P.A. at 90.5, 102.5, 104.5 GHz. The upper limit at 92.5 GHz with a signal-to-noise ratio of 3 is also plotted with a red arrow. The grey dotted line indicates the constant fitting function with a reduced chi-square of 4.5 (d.o.f = 3).}
\label{rotation}
\end{figure*}

\subsection{Polarization}

Figure \ref{polimage}a shows the Stokes $I, Q,$ and $U$ maps obtained
using the entire
ALMA Band 3 frequency range taken 5.2 days after the GRB. Detections
with a confidence level of 5$\sigma$ or better on the $Q$ and $U$ maps
yield a polarization degree of $0.27 \pm 0.04\%$ (including systematic error). 
Our measured value describes the intrinsic origin because
depolarization between the source and observing site is negligible for
the point source (i.e. GRB afterglows) in this millimeter
band \citep{24}. Although we could not find any detection in the Stokes
$Q$ and $U$
maps at 11.2 days, we measured the corresponding deep upper limit of
the polarization degree ($<0.27\%$, 3$\sigma$ significance), which was
consistent with that at 5.2 days within the error margin.

The apparent brightness of $31.94 \pm 0.44~$mJy observed 5.2 days after
the burst using the entire ALMA Band 3 frequency range enabled more
detailed polarimetric analysis using four individual spectral windows
(SPW) of Band 3 (Figure \ref{polimage} b, c, d, and e). The measurments are summarized in Table \ref{almapol}. Other than the Stokes $U$ map
at 92.5 GHz, there were significant detections at a 3.0$\sigma$
confidence level or better. In the Stokes $U$ map at 92.5 GHz, there was
no significant flux, and the range of the
P.A. was constrained. Although the polarization degrees in each SPW
were consistent with the value measured using the entire Band 3
frequency, the P.A. significantly varied with the wavelength (Figure
\ref{rotation}). The observed P.A. is most likely intrinsic value because the
Faraday rotation effect for both the host galaxy and Milky Way Galaxy
is quite small at this frequency \citep{26,27}. The expected galactic
Faraday rotation effect is up to $\sim0.3^{\circ}$.
We tried to fit the P.A. data including the upper limit \citep[the method is described in][]{28} with
constant or linear function of squared wavelength, but did not obtain a good fit (Figure \ref{rotation}).

\section{Discussion} \label{sec:discuss}

\subsection{Afterglow modeling}

%The spectral data and the light curve at 230 GHz for 1$-$10 days can
%be explained by the standard forward shock model \citep[e.g.][]{22}.
%The closure relation \citep{33} at $t\gtrsim4$ days is
%$\alpha-3\beta/2 \simeq -0.69\pm0.07$.
%The standard synchrotron
%emission from the forward shock expanding in uniform density medium
%for $\nu_{m} < \nu < \nu_{c}$ obeys $\alpha-3\beta/2$ = 0 when the
%edge of the collimated shock is not observed due to the relativistic
%beaming effect. After the edge can be observed (but the shock does not
%expand sideways), the additional geometrical flux reduction
%$\Gamma^{2}\theta_{j}^{2} \propto t^{-3/4}$ leads to $\alpha -
%3\beta/2 = -3/4$, where $\Gamma$ and $\theta_{j}$ are the Lorentz
%factor and opening half-angle of the shock, respectively. The latter
%relation is consistent with our observation. The decay index for
%$\nu_{a} < \nu < \nu_{m}$ in this model is $\alpha=-1/4$, which is
%also consistent with that for $t\lesssim4$ days. Therefore, the break
%in the 230 GHz light curve is due to the $\nu_{m}$ crossing.
We find that the spectra of Figure \ref{polspec}a can be well fitted by the foward shock synchrotron emission model by \citet{22} with the synchrotron self-absorption frequency $\nu_a \sim 20\;$GHz and the synchrotron frequency of minimum-energy electrons $\nu_m \sim 200\;$GHz. In such a late phase the slow cooling regime (i.e. $\nu_m < \nu_c$) is likely.
Then the observed shallow decay at $t \lesssim 4\;$days may correspond to the spectral segment $\nu_a < \nu < \nu_m$.
If the spectrum in this segment is the power-law with $\beta=1/3$, the decay index is $\alpha = 3\beta/2-1/2 = 0$ in the wind environment case \citep{33}. However, our smoothly broken power-law spectrum ($\beta<1/3$ effectively, see Figure \ref{polspec}a) leads to a steeper decay, which cannot fit the observed $230\;$GHz light curve at $t \lesssim 4\;$days. Alternatively, we find that the ISM environment case can well fit it.

The flux after $\nu_m$ crosses the observed frequency (i.e. $\nu_m < \nu < \nu_c$) obeys the closure relation $\alpha - 3\beta/2 = 0$ \citep{33} in the ISM environment case when the edge of the collimated shock is not observed. After the edge is observed \citep[but the shock does not expand sideways;][]{zhang}, the additional geometrical flux reduction $\Gamma^{2}\theta_{j}^{2} \propto t^{-3/4}$ leads to $\alpha - 3\beta/2 = -3/4$, where $\Gamma$ and $\theta_{j}$ are the Lorentz factor and opening half-angle of the shock, respectively. The latter relation is consistent with the observed relation $\alpha - 3\beta/2 \simeq -0.69 \pm 0.07$.

%Based on this model, the spectra at $t = 4.3$ and 5.2 days and the
%multiband light curves can be well fitted by
Based on the above consideration, we adopt 
the flux formula of
\citet{22} multiplied by the geometrical flux reduction factor
$[1+(t/t_{j})]^{-3/4}$ to fit the observed data
(Figure \ref{lc}, \ref{polspec}a).
Here we set the
synchrotron self-absorption frequency $\nu_{a}\simeq 22~$GHz, the
synchrotron frequency of minimum-energy electrons
$\nu_{m} \simeq 200\;(t/4.3~{\rm days})^{-3/2}$ GHz, the peak flux before the
jet break $F_{\nu_{m}} (t<t_{j})\simeq 72~$mJy, the jet break time
$t_{j} \simeq 2~$days, and the electron energy spectral index
$p \simeq 3$. The first three characteristic quantities are functions of
four physical parameters, namely the isotropic shock energy $E_{\rm iso}$,
the ambient medium density $n$, the fraction of shock energy carried
by the electrons $\epsilon_{e}$, and that carried by amplified
magnetic field $\epsilon_{B}$.
Thus, we have the relations $n \simeq 600\;(E_{\rm iso}/5 \times 10^{48}~{\rm erg})^3~{\rm cm}^{-3}$,
$\epsilon_e \simeq 0.3\;(E_{\rm iso}/5 \times 10^{48}~{\rm erg})$,
and $\epsilon_B \simeq 0.2\;(E_{\rm iso}/5 \times 10^{48}~{\rm erg})^{-5}$.
The numerical values of $n, \epsilon_e,$ and $\epsilon_B$ are not unrealistic \citep{23}, and $E_{\rm iso}$ should not be considerably different from this value because of $\epsilon_e < 1$ and $\epsilon_B < 1$.
%\footnote{Our model is required to have inefficient electron acceleration at high energies, i.e. $p \gtrsim 4$, for not overwhelming the observed flux in the X-ray band at $t \sim 5000\;$s. The observed X-ray light curve has a large variability, which we consider as a different origin.}
For these values we calculated X-ray light curve, which does not overwhelm the observed one.
This analysis means that we found a possible physical afterglow model (while we leave a full exploration of possible models to separate work), and supports our argument that
we performed the first radio afterglow
polarimetry in the waveband well above $\nu_a$ \citep[c.f.][]{40,42}.

\subsection{Faraday depolarization effect}

We focused on the polarization at 5.2 days, the phase when the
intensity can be explained by the standard forward shock model.
The precise detection of the polarization degree of $0.27 \pm 0.04\%$
indicated that the value is the smallest one among all afterglow
polarization measurements, and smaller than those in late-time optical
afterglows explained by the standard forward shock model, which range
from 0.5\% to 10\% \citep{10,11,12} 
\footnote{Although the minimum value of 0.31\% was measured with the GRB030329 optical afterglow, the measurment was performed during the multiple bump light curve phase with strong polarization variabilities (i.e. extra physical explanations to the standard afterglow model are required).}.

There was no polarimetric data at the higher frequency ranges for the
present event (except the supernova component in the optical band).
%we assumed $P_{0}=1\%$.
%%%%%
% optical polarization
%%%%%
Note that there are 84 polarimetric measurements for optical
afterglows (i.e. excluding measurements for early-time reverse shock
components that show high values) among 13 GRBs \citep{12}. The
weighted average and average of the measurments are $1.0\%$ and $1.6\%$,
respectively. Among these, 58 measurements were made during the phases
in which the intensities are describable by the standard forward shock
model. For these selected events, the weighted average and average of
the linear polarizations are $1.2\%$ and $1.7\%$, respectively.

By assuming a polarization degree at higher frequency ranges
(e.g. optical) for the present event as $P_0 = 1\%$, we calculate the
polarization spectrum based on the afterglow model described above
\citep[c.f.][]{34,25,37,38}, and plot it by the green dotted line in
Figure \ref{polspec}b.  It varies by a factor of $0.5(p+7/3)/(p+1)
\simeq 2/3$ at $\nu = \nu_m$ and decays at $\nu \lesssim \nu_a$.  Our
measured value is substantially lower than this model line.

If only part of the swept-up electrons is accelerated, the
non-accelerated electrons with thermal Lorentz factor
$\tilde{\gamma}_m = \eta\Gamma$ cause Faraday depolarization at
$\nu>\nu_{a}$ \citep{9}, where $\eta$ is a factor of the order of
unity in the case that the non-accelerated electrons are just
isotropized at the shock front \citep{6}. Such a model in which the
fraction of accelerated electrons is $f < 1$ can explain the intensity
in the same way as in the standard model with the parameters $E_{\rm
  iso}'$ = $E_{\rm iso}/ f$, $n'=n/f$, $\epsilon_{e}'=\epsilon_{e}f$,
and $\epsilon_{B}'=\epsilon_{B}f$ \citep{6}.
Thus, a very small value
of $f$ would lead to a crisis of the total energy requirement.  In
this scenario, the polarization degree is given by $P_0
\sin(\tilde{\tau}_V/2)/(\tilde{\tau}_V/2)$ where $\tilde{\tau}_V =
(\nu/\tilde{\nu}_V)^{-2}$ and $\tilde{\nu}_V \sim
200\;[(1-f)/10f]^{1/2} \eta^{-1} \sqrt{\ln \tilde{\gamma}_m} N^{-1/12}
(E_{\rm iso}/10^{52}\;{\rm erg})^{3/16} \\ n^{9/16}
(\epsilon_B/0.01)^{1/4} (t/1\;{\rm day})^{-1/16}\;$GHz.
Here the
magnetic field in the shocked region has been assumed to be tangled on
hydrodynamic scales, following \cite{9} and \cite{41}, and then the
plasma can be considered to consist of a number of random cells, in
each of which magnetic field is ordered \citep{37,39}. $N$ denotes the
number of random cells in the three-dimensional visible region. In
this case $P_0 = (p+1)/[(p+7/3)\sqrt{N}]$ for $\nu > \nu_m$ while $P_0
= 0.5/\sqrt{N}$ for $\nu_a < \nu < \nu_m$.  With $P_{0}=1\%$ for $\nu
> \nu_m$, $\tilde{\nu}_V \simeq 210~$GHz explains our measurement (see
the green dashed line in Figure \ref{polspec}b), which corresponds to
$1/f \sim 12\;(E_{\rm iso}/2\times10^{50} {\rm erg})^{-5/4}
\eta^{2}(\ln \tilde{\gamma}_m)^{-1}$. For the case of $P_{0} = 0.5\%$
(Figure \ref{polspec}b), $1/f \sim 10$ is still required.
For the case of $P_0 = 10\%$, $1/f \sim 60$ is required.
We should also note that the case of $P_0 \simeq 0.33\%$ is not ruled out (see Figure \ref{polspec}b), where the Faraday depolarization effect with $f<1$ is not required.

The P.A. becomes a very complicated function of wavelength and the
functional form is determined randomly for such a tangled magnetic
field that we assume\citep{26}.  Therefore, the observed variation of
the P.A. is not inconsistent with this scenario.

In summary, with the first intensive combined use of telescopes in the
millimeter and submillimeter ranges for the GRB171205A afterglow, our
observations provided the first linear polarimetry in the millimeter
band. The measured polarization degree is substantially lower than the typical optical
one. Although the (semi-) simulataneous measurments in multiple
wavelengths are required, this measurment suggests the Faraday
depolarization effect and larger total energy by a factor of $\sim 10$
than ordinary estimate without considering non-accelerated
electrons. The observed P.A. variation along with wavelength
is not inconsistent with this scenario.
Multi-frequency
polarimetry in the submillimeter/millimeter range and/or with
simultaneous optical polarimetry would provide more accurate
non-accelerated electron fraction. Hence, this observation consolidates
the new methodology for revealing the fundamental properties of GRBs.

\begin{table*}[hbtp]
%\captionsetup{labelformat=empty,labelsep=none} 
\caption{Polarization and Photometric Observing Log}
\label{almapol}
\centering
\begin{tabular}{ccccccc}
\tableline
\multicolumn{7}{l}{Polarimetry Epoch1: 2017-12-10 10:23-13:17, T=5.187 days }\\ 
SPW & Frequency [GHz] & Pol. [\%] & P.A. [deg] & I flux [mJy] & Q flux [mJy] & U flux [mJy] \\
\tableline
0,1,2,3 &  97.5 & 0.27$\pm$0.04 & $-$71.3$\pm$3.3     & 31.944$\pm$0.440 & $-$0.069$\pm$0.009 & $-$0.053$\pm$0.011\\
0   &  90.5 & 0.30$\pm$0.06 & $-$67.9$\pm$4.7     & 32.719$\pm$0.413 & $-$0.070$\pm$0.010 & $-$0.069$\pm$0.020\\
1   &  92.4 & $<$0.32        & $<-$78.1 or $>$78.1 & 32.514$\pm$0.365 & $-$0.094$\pm$0.026 & 0.014(rms) \\
2   & 102.5 & 0.35$\pm$0.08 & $-$71.3$\pm$5.5     & 31.172$\pm$0.399 & $-$0.086$\pm$0.025 & $-$0.066$\pm$0.018\\
3   & 104.5 & 0.31$\pm$0.06 & $-$58.0$\pm$4.9     & 30.898$\pm$0.412 & $-$0.043$\pm$0.012 & $-$0.087$\pm$0.029\\
\tableline
\multicolumn{7}{l}{Polarimetry Epoch2: 2017-12-16 11:14-14:33, T=11.231 days}\\ 
SPW & Frequency [GHz] & Pol. [\%] & P.A. [deg] & I flux [mJy] & Q flux [mJy] & U flux [mJy] \\
\tableline
All &  97.5 & $<$0.27 & --- & 15.705$\pm$0.090 & 0.010 (rms) & 0.010 (rms) \\ 
0   &  90.5 & $<$0.52 & --- & 16.171$\pm$0.106 & 0.020 (rms) & 0.020 (rms) \\ 
1   &  92.4 & $<$0.52 & --- & 16.054$\pm$0.110 & 0.019 (rms) & 0.019 (rms) \\ 
2   & 102.5 & $<$0.52 & --- & 15.370$\pm$0.113 & 0.019 (rms) & 0.019 (rms) \\ 
3   & 104.5 & $<$0.54 & --- & 15.206$\pm$0.111 & 0.019 (rms) & 0.019 (rms) \\ 
\tableline
\multicolumn{7}{l}{Total Flux Observation Log}\\
Instrument & Epoch [days] & Frequency [GHz] & Flux [mJy] & \multicolumn{3}{l}{}\\
\tableline
VLA & 4.306 &  5.0 &  2.41$\pm$0.12 & \multicolumn{3}{l}{}\\
VLA & 4.306 &  7.1 &  4.32$\pm$0.05 & \multicolumn{3}{l}{}\\
VLA & 4.306 &  8.5 &  5.71$\pm$0.05 & \multicolumn{3}{l}{}\\
VLA & 4.306 & 11.0 &  8.42$\pm$0.06 & \multicolumn{3}{l}{}\\
VLA & 4.306 & 13.5 & 11.26$\pm$0.09 & \multicolumn{3}{l}{}\\
VLA & 4.306 & 16.0 & 14.01$\pm$0.11 & \multicolumn{3}{l}{}\\
SMA &  1.496 & 230 & 53.6$\pm$0.9 & \multicolumn{3}{l}{}\\
SMA &  2.412 & 230 & 48.4$\pm$0.6 & \multicolumn{3}{l}{}\\
SMA &  3.478 & 230 & 41.2$\pm$0.8 & \multicolumn{3}{l}{}\\
SMA &  4.272 & 230 & 30.0$\pm$0.7 & \multicolumn{3}{l}{}\\
SMA &  8.398 & 230 & 11.1$\pm$1.0 & \multicolumn{3}{l}{}\\
SMA & 11.033 & 230 &  8.9$\pm$0.4 & \multicolumn{3}{l}{}\\
SMA &  3.478 & 345 & 21.9$\pm$2.3 & \multicolumn{3}{l}{}\\
ACA &  5.126 & 345 & 17.0$\pm$0.8 & \multicolumn{3}{l}{}\\
ACA &  7.069 & 345 & 12.9$\pm$0.1 & \multicolumn{3}{l}{}\\
SMA &  8.398 & 345 & $>$15.8      & \multicolumn{3}{l}{}\\
ACA & 11.180 & 345 &  6.9$\pm$0.3 & \multicolumn{3}{l}{}\\ 
\tableline
\end{tabular}
\end{table*}

\acknowledgments

We thank the anonymous referee for his/her careful review of our manuscript.
This paper makes use of the following ALMA data:
ADS/JAO.ALMA\#2017.1.00801.T. ALMA is a partnership of ESO
(representing its member states), NSF (USA), and NINS (Japan),
together with NRC (Canada), MOST and ASIAA (Taiwan), and KASI
(Republic of Korea), in cooperation with the Republic of Chile. The
Joint ALMA Observatory is operated by ESO, AUI/NRAO, and NAOJ. This
work is supported by the Ministry of Science and Technology of Taiwan
grants MOST 105-2112-M-008-013-MY3 (Y.U.) and 106-2119-M-001-027
(K.A.). This work is also supported by JSPS Grants-in-Aid for
Scientific Research No. 18H01245 (K.T.).
We thank EA-ARC, especially Pei-Ying Hsieh for
support in the ALMA observations. We also thank P. T. P. Ho and Y. Ohira for helpful comments. Y.U, K. Y. H, and K. A. also thank Ministry of
Education Republic of China.

%% To help institutions obtain information on the effectiveness of their 
%% telescopes the AAS Journals has created a group of keywords for telescope 
%% facilities.
%
%% Following the acknowledgments section, use the following syntax and the
%% \facility{} or \facilities{} macros to list the keywords of facilities used 
%% in the research for the paper.  Each keyword is check against the master 
%% list during copy editing.  Individual instruments can be provided in 
%% parentheses, after the keyword, but they are not verified.

\vspace{5mm}
\facilities{ALMA, SMA, VLA}

%% Similar to \facility{}, there is the optional \software command to allow 
%% authors a place to specify which programs were used during the creation of 
%% the manusscript. Authors should list each code and include either a
%% citation or url to the code inside ()s when available.

\software{CASA \citep{32},  
          %Cloudy \citep{2013RMxAA..49..137F}, 
          %SExtractor \citep{1996A&AS..117..393B}
          }

\end{document}